# Formal Modeling in Software Engineering with LLMs: A Systematic Literature Review

Alexandra Newcomb and Omar Ochoa

***Abstract*—Formal verification provides strong correctness guarantees for software systems but remains hindered by the substantial expertise and manual effort required to construct and maintain formal models. The recent rise of Large Language Models (LLMs) has prompted growing interest in whether their natural language understanding and generation capabilities can be leveraged to automate labor-intensive modeling activities. This paper presents a Systematic Literature Review (SLR) examining how LLMs have been applied to automate model checking and the generation of formal and semi-formal models within software engineering. The review follows established SLR methodologies and a PICOC-driven search strategy across four major digital libraries. From an initial search of 1,193 articles, 34 peer-reviewed conference, journal, and workshop articles published between 2022 and April 2026 were selected and analyzed using a structured data extraction form. The review identifies six recurring trends in the literature, including the automated generation of formal models for model checking, LLM-assisted downstream formal verification, iterative and repair-based generation pipelines, and the generation of semi-formal UML and SysML artifacts whose underlying techniques are transferable to formal modeling contexts. Across these trends, LLMs are generally found to perform well on syntactic generation and repair tasks but struggle with semantic correctness, motivating widespread use of counterexample-guided repair loops, intermediate representations, and human-in-the-loop validation. Notably, this review finds that the use of LLMs specifically to automate model checking remains a largely unexplored area, with comparatively little research published on this narrower task. By synthesizing the common methods, trends, and limitations of this emerging body of work, this SLR provides researchers entering this topic with a structured overview of current progress. This SLR also offers a roadmap of open gaps and promising directions to guide future contributions at the intersection of LLMs and formal methods.**

This work is supported by the National Science Foundation under Grant No. 2445056. Any opinions, findings, and conclusions or recommendations expressed in this material are those of the authors and do not necessarily reflect the views of the National Science Foundation.

Alexandra Newcomb is a Ph.D. student at Embry-Riddle Aeronautical University in Daytona Beach, FL 32114 USA (e-mail: davidofa@my.erau.edu).

Omar Ochoa is an Associate Professor of Software Engineering and Computer Science at Embry-Riddle Aeronautical University in Daytona Beach, FL 32114 USA (e-mail: ochoao@erau.edu).

## I. INTRODUCTION

THE rapid advancement of Large Language Models (LLMs) is reshaping how software is designed, implemented, and verified [1]. Researchers are increasingly exploring whether the Natural Language (NL) understanding and generation capabilities of these models can be further extended to rigorous and expertise-intensive practices within software engineering. Among these practices, formal verification, and model checking in particular, stands out as areas with significant potential to benefit from LLM-based automation, given their historically high barrier to entry and substantial manual effort requirements [2, 3]. This paper presents a Systematic Literature Review (SLR) examining how LLMs are currently applied to automate model checking and the generation of formal models within software engineering.

Model checking is a widely used formal verification technique that exhaustively explores the state space of a system model to determine whether the system satisfies a specified property, such as safety or liveness properties [4, 5]. Because model checking provides strong correctness guarantees that cannot be obtained through testing alone, it has been applied across numerous software engineering domains, including reactive systems, security protocols, business processes, and safety-critical systems [4, 6, 7, 8]. Despite these benefits, model checking and formal verification more broadly continue to face significant barriers to industrial adoption. First, constructing a formal model or specification that accurately captures a system's requirements and behavior requires substantial expertise in formal methods. Second, even for experienced modelers, the process of manually authoring, refining, and verifying formal models and their associated properties is time-consuming, which discourages their use in fast-paced development environments [2, 3, 9]. These two challenges limit the widespread adoption of formal methods in industry.

In parallel, LLMs have been increasingly adopted to automate a wide range of software engineering activities. LLMs have been used to automatically generate source code from NL descriptions, to generate test cases, and to assist with program repair and code review, among other tasks [10, 11, 12, 13]. Building on this momentum, researchers have begun to investigate whether LLMs can similarly automate aspects of formal methods. One especially active line of work is formal requirements specification, in which an LLM is used to translate NL requirements into formal representations, such as temporal logic formulas, that can subsequently be used for formal analysis [14, 15, 16]. This emerging body of work suggests that LLMs, given their demonstrated ability to bridge NL and formal notations, may be similarly capable of automating the construction of formal models themselves. By extension, LLMs may ease the adoption of model checking and other formal verification.

TABLE I
PICOC TERMS

| PICOC Criteria | Term | Synonyms |
|---|---|---|
| Population | Model Checking | Formal Model |
| Intervention | Large Language Models | N/A |
| Comparison | N/A | N/A |
| Outcome | Automation | Efficiency, Usability, Ease of Use, Scalability |
| Context | Software Engineering | Software Verification |

This work is specifically interested in how LLMs can assist software engineers in formal modeling for model checking and other downstream formal verification activities that require the construction of a formal system model. Because the literature directly addressing this area remains relatively sparse, this SLR is secondarily interested in how LLMs have been used to generate semi-formal representations of system behavior outside the context of formal verification, such as through the generation of Unified Modeling Language (UML) and System Modeling Language (SysML) diagrams. Articles addressing such semi-formal modeling tasks are therefore included in this review, as the underlying techniques used to generate these artifacts, including prompting strategies, iterative repair mechanisms, and human-in-the-loop refinement, may be readily transferable or adaptable to the generation of fully formal models intended for model checking.

This work provides a detailed synthesis of the common trends, methods, and ideas shared across the selected studies, as well as the gaps, challenges, and areas for future work that emerge from this body of research. As a result, this SLR serves as a primer for researchers seeking to enter this area of study by offering a structured overview of how LLMs have thus far been applied to formal and semi-formal software modeling. At the same time, by identifying the gaps and limitations present in current approaches, this review offers a roadmap to guide future contributions to the field.

Ultimately, continued research at the intersection of LLMs and formal modeling has the potential to meaningfully increase the adoption of formal methods in industry. LLMs are well-positioned to lower the barrier to entry for formal verification by improving the understandability of formal notations and by reducing the manual effort historically required to construct and maintain formal models. Research in this area also increases the accessibility of model checking to software engineers who lack specialized formal methods training.

## II. RESEARCH METHOD

This section details the methods used to conduct the SLR. The method is adapted from methodologies presented by Carrera-Rivera et al., Kitchenham and Charters, and Claes [17, 18, 19].

### *A. PICOC and Synonyms*

When framing the research questions and search terms, the Population, Intervention, Comparison, Outcome, and Context (PICOC) method was used [17, 18]. Table I presents the PICOC terms for this research. These terms are used to form the research questions and search strings presented in sections II.B and II.D respectively.

### *B. Research Questions*

This systematic review addresses the following three Research Questions (RQs), formulated from the PICOC terms.

1) How are researchers applying LLMs to automate model checking or generate formal and semi-formal models for software engineering activities?
2) What common ideas or methods do the selected works from RQ1 share?
3) What gaps, challenges, and areas for future work exist within the current literature applying LLMs to model checking or formal model generation?

### *C. Digital Library Sources*

Articles for review were selected by querying the databases below with the search terms presented in section II.D. These databases were selected due to their relevance to this area and reputability for producing high-quality research.

1) IEEE Xplore
2) ACM Digital Library
3) Springer
4) Science Direct by Elsevier

Only peer-reviewed conference, journal, and workshop articles were included. After the above four databases were individually queried, an independent search through the first 200 results in Google Scholar on the search string was conducted to obtain any additional articles that may have been missed by the databases and that belonged to one of the four databases.

Additionally, backward snowballing was used to add any articles referenced in the selected articles that did not appear in the original search yet were supported by the inclusion criteria [19]. Articles found through snowballing also originated from the selected digital libraries.

### *D. Search Strings*

Using the PICOC terms, the following search string was synthesized: ("Model Checking" OR "Formal Model") AND "Large Language Models" AND ("Software Engineering" OR "Software Verification").

Note that the Outcome terms from the PICOC terms were excluded from the search string to avoid unnecessarily limiting the search results. Articles were instead filtered to relate to the Outcome as pertains to RQ1.

TABLE II
INCLUSION AND EXCLUSION CRITERIA

| Criteria | Inclusion | Exclusion |
|---|---|---|
| Time Period | 2020 to April 2026. | Articles published before 2020 and after April 2026. |
| Language | Articles written in English. | All articles not written in English. |
| Type of Literature/Source | Conference, Workshop, and Journal articles presenting original empirical or theoretical research. | All other articles, including grey literature and preprints. Articles that do not present empirical or theoretical research. |
| Relevance to Research Questions | Relevant articles to RQ1. | Articles irrelevant to RQ1. |

### *E. Inclusion and Exclusion Criteria*

The inclusion and exclusion criteria used for the SLR are presented in Table II. Criteria are adapted from Carrera-Rivera et al. [17]. Due to the limited nature of work in this area, workshop articles were included despite their lower quality when compared to conference and journal articles. Workshop articles frequently present novel ideas for future exploration. Additionally, articles were chosen no earlier than 2020 due to the significant amount of research in ML over the last six years. Note that for ACM Digital Library, the time period was chosen to be from 2021 to 2026, as the ACM search engine did not permit a start date earlier than 2021.

Although the primary focus of this SLR is on automating model checking and formal modeling for downstream formal verification, articles focusing on semi-formal modeling are included due to the relatively low number of articles originating in the former area. These articles remain applicable to formal modeling, as methods detailing the automation of semi-formal modeling through LLMs can be adapted to more formal contexts. However, the focus on formal modeling and the use of the search term presented in section II.D may have resulted in the exclusion of some articles on UML and SysML modeling that do not contain those search terms. A complete focus on UML and SysML modeling through LLMs would steer the scope of this review largely away from formal modeling, which is why this potential exclusion does not necessarily reduce the rigor of the methodology. The primary focus remains on formal modeling.

### *F. Data Extraction Form*

After collecting relevant articles meeting the inclusion and exclusion criteria, data was systematically collected using the data extraction form presented in Table III. The data extraction form was curated following guidelines presented by Kitchenham and Charters [18]. The form was designed to collect bibliometric information about each article and to answer the RQs.

### *G. Data Analysis and Synthesis*

Once all articles were obtained using the search terms, inclusion criteria and exclusion criteria, and data was extracted through the data extraction form, data analysis was performed. Data analysis and synthesis occurred by analyzing both bibliometric information and technical information. Bibliometric results were synthesized, such as the digital library database producing the largest number of articles in this area, the distribution of literature types, and the countries where research in this area is being conducted. Similarly, technical results were gleaned from the literature by comparing technical data across all studies to answer the RQs. Sections III and IV present the bibliometric and technical results of the SLR respectively. Section V answers RQ2, and section VI answers RQ3.

TABLE III
DATA EXTRACTION FORM

| Data Item | Value |
|---|---|
| | *Bibliographic Information* |
| Title | The full title of the research article. |
| Authors | Follows in-text citation guidelines. |
| Database source | Ex. “IEEE Xplore” |
| Literature type | Ex. “Journal” |
| Name of conference | Ex. “N/A |
| Name of journal | Ex. “IEEE Robotics and Automation Letters” |
| Name of workshop | Ex. “N/A” |
| Country of origin | Fully spelled out (i.e. no abbreviations). |
| Year | Ex. “2026” |
| | *Technical Information* |
| Research questions or aim, goals, problem, objectives, motivation, purpose | This field provides background and a rationale for the empirical research presented in the article. |
| LLM model or framework used | Ex. “GPT-4o and GPT-5-mini” |
| Model checking or formal modeling method used | Ex. “Storm probabilistic model checker. The models are eventually translated into the PRISM specification language.” |
| Application of LLM to Model Checking or formal modeling | A brief description of the solution or framework the authors propose, including all key elements. |
| Summary of experimental design | A brief description of the experiment used to evaluate the proposed solution or framework. |
| Summary of experimental results | A brief summary of the quantitative and qualitative results of the research, noting if findings are credible and useful. |
| Key findings and analysis | A summary of the key findings of the research article. |
| Limitations, challenges, threats to validity | Either directly stated by the authors or inferred from the research. |
| Identified future work | Directly stated by the authors. |
| Additional notes | Optional: any notable remarks about the article. |

## III. BIBLIOMETRIC RESULTS

The initial search resulted in 1,193 articles, of which 528 originated from IEEE Xplore, 302 originated from ACM Digital Library, 79 originated from Springer, 84 originated from Science Direct by Elsevier, and the remaining 200 articles originated from Google Scholar. The initial search already excluded articles before 2020, articles not in English, and that were not of the allowed literature type. By reading through titles and abstracts, a total of 34 articles were selected across a variety of conferences, journals, and workshops. Seven articles were found through backward snowballing, a process described in section II.C.

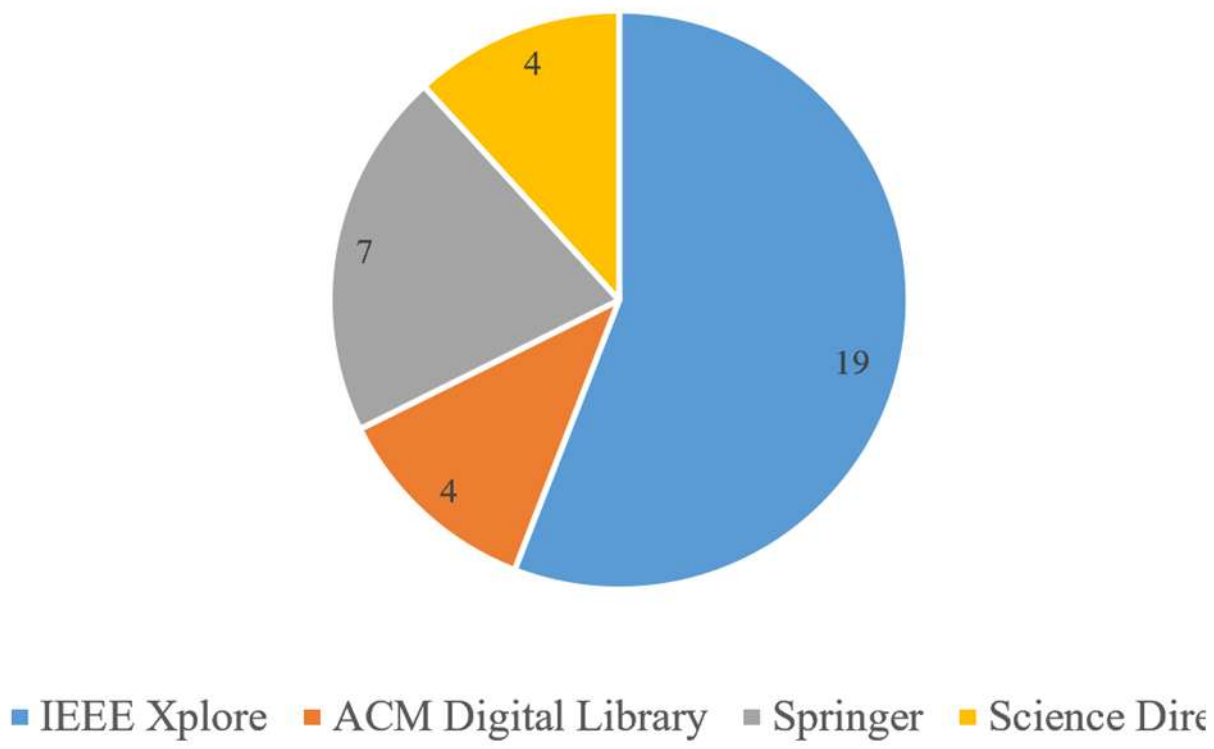

**Fig. 1.** Distribution of articles across selected databases. Values represent the count of articles belonging to that category.

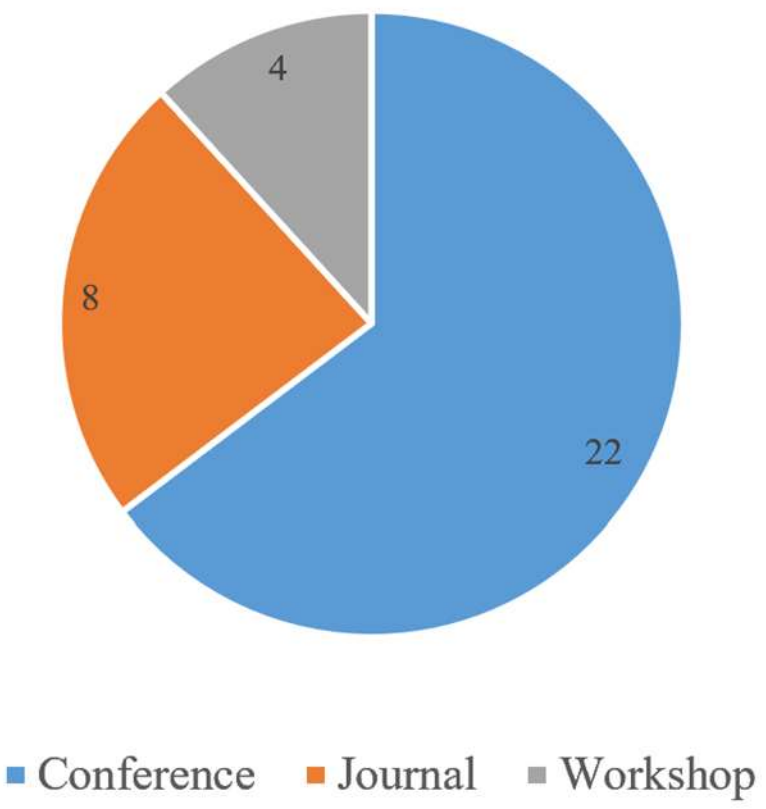

**Fig. 2.** Distribution of selected articles across article type (conference, journal, or workshop). Values represent the count of articles belonging to that

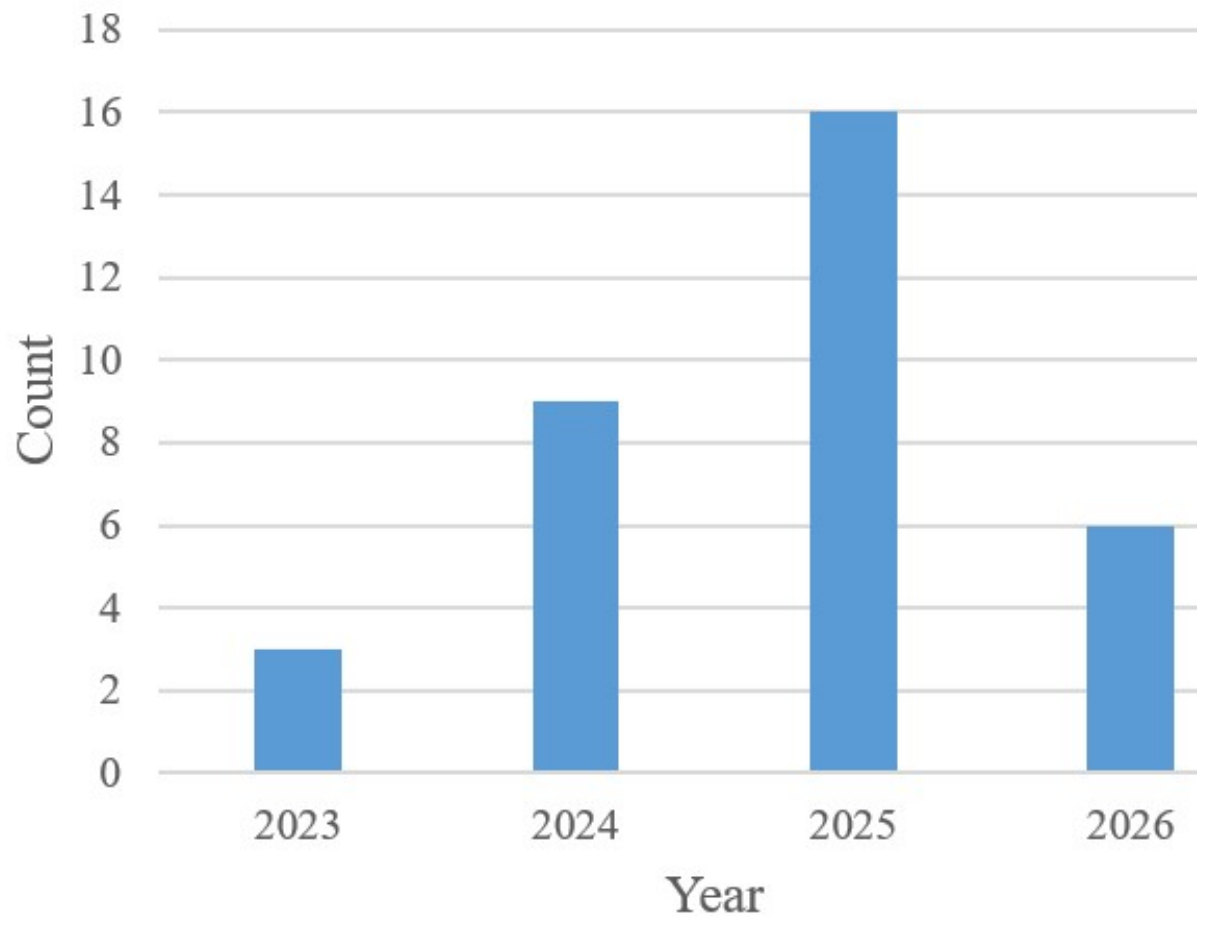

**Fig. 3.** Distribution of articles by year.

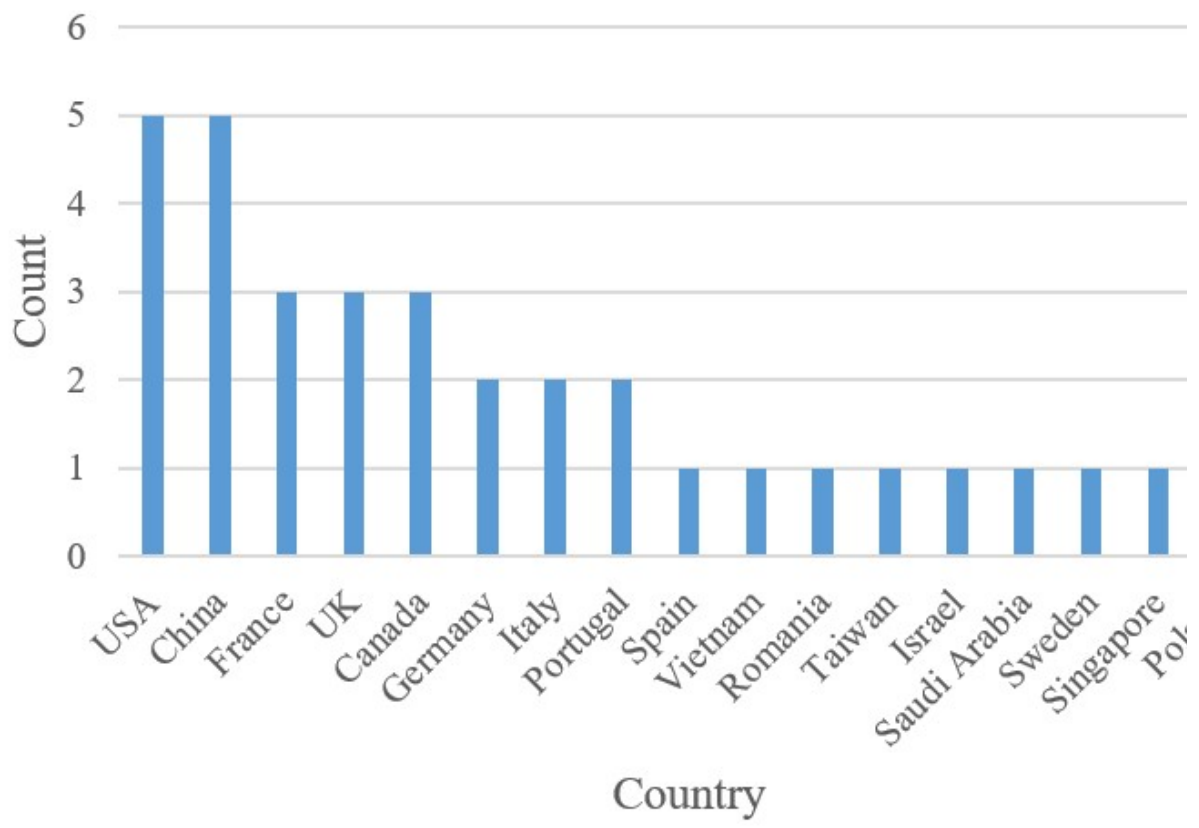

**Fig. 4.** Distribution of articles by country. United States of America is abbreviated as 'USA'. Similarly, the United Kingdom is abbreviated as 'UK'.

Fig. 1 presents the distribution of articles across the included databases. Notably, most articles (56%) originated from IEEE Xplore, with the other three databases producing significantly less research in this area. Conference papers originating from IEEE/ACM joint conferences were generally listed under IEEE Xplore, with the exception of one article that appeared in ACM Digital Library but not IEEE Xplore. Similarly, Fig. 2 presents the number of conference, journal, and workshop articles that resulted from the literature review. Most selected articles were conference papers (65%), with significantly fewer journal articles and workshop articles. Most research in this area is disseminated via conferences, especially IEEE or ACM conferences.

Figs. 3 and 4 present the distribution of articles over time and the distribution of articles by country. Although the inclusion criteria included articles after 2020, because LLMs were largely unavailable before 2022, the selected articles all originated after 2022. Most articles were published in 2025, with a notable number of articles already published in 2026 before the April 2026 cutoff. The country of origin was determined by viewing the country that the first author is affiliated with for each article. Notably, articles originated from a diverse set of 17 countries, with the largest proportion of articles originating from the United States and China.

The selected articles were disseminated across a diverse list of conferences, journals, and workshops. Two conferences yielded the largest number of articles for this literature review: IEEE/ACM International Conference on Automated Software Engineering (ASE) and IEEE/ACM International Conference on Model Driven Engineering Languages and Systems (MODELS). Both conferences yielded 3 papers each. Otherwise, three conferences yielded two papers each (from IEEE and ACM), and the other conference papers all originated from distinct conferences. All journal articles originated from distinct journals. Notably, two workshop papers (out of 4) originated from the IEEE/ACM International Conference on Automated Software Engineering - Workshops (ASEW), a workshop extension of ASE.

## IV. Technical Results

This section answers RQ1 and identifies six key trends in how the literature utilizes LLMs to automate model checking or generate formal models for software engineering activities.

The defined trends are not mutually exclusive, and articles may be discussed in multiple sections. The six identified trends are as follows and discussed in respective sections below:

1) Automating Model Checking
2) Formal Models for Downstream Formal Verification
3) Iterative and Repair-Based Methods
4) UML Model Generation
5) SysML Model Generation
6) Generating Models from Code

### *A. Automating Model Checking*

Two trends were discovered in the area of automating model checking. Half of the articles focused on automating the generation of formal models for model checking, while the other articles discussed other forms of automating model checking. Examples of the latter include repairing specifications, generating formal properties from formal models, and generating understandable descriptions of counterexamples [20, 21, 22].

#### *1) Automating the Generation of Formal Models for Model Checking*

Seven articles focus on automating the generation of formal models for model checking using LLMs [23, 24, 25, 26, 27, 28, 29]. Saccon et al. automatically synthesize Markov Decision Processes (MDPs) using LLMs [23]. They then use the Storm model checker to determine an optimal policy over the generated MDP based on a specified property. This optimal policy can then be exported as a state-action table and deployed by a robotic system. Policies are automatically generated from NL descriptions of the scenario. The authors found that the LLMs (GPT-4o and GPT-5-mini in their experiments) were able to generalize across multiple domains [23]. Notably, this is the only work discovered in the literature review that automatically synthesized MDPs for model checking using LLMs, indicating that this area is largely unexplored.

Ramani et al. evaluate how well LLMs take NL plans as input and generate corresponding Kripke structures, which can then be model checked [24]. They specifically perform Bounded Model Checking (BMC) through NuSMV. They also automatically formalize the LTL requirement using LLMs. Therefore, the LLM framework generates both the model (Kripke structure) and the property (LTL). The authors found that GPT-5 achieves high performance on syntactically generating the formal models from NL plans. On the other hand, GPT-4o achieved significantly poorer performance [24].

Similarly, Jha et al. investigate how well LLMs can co-synthesize code, a formal model of the code, and a functor that maps elements from the code to the model from NL descriptions [25]. The aim is to reduce the manual effort required to verify code, especially code with concurrency and synchronization. From the NL problem description, the LLM concurrently generates C code, a formal model in SMV (to be checked by the NuSMV model checker), and a functor. This article is the only article in the literature review that discusses the use and synthesis of a functor, which serves as a human-auditable artifact that maps elements from the code to elements in the formal model. The generation of the functor potentially increases the probability that the generated formal model correctly captures the generated code and makes both components more understandable to the human developer [25].

Zuo et al., Nivon et al., and Sultan and Apvrille present user-facing tools for generating formal models [26, 27, 28]. Zuo et al. propose PAT-Agent, an end-to-end LLM-based framework for NL auto-formalization and formal model repair [26]. A planning LLM extracts modeling elements such as constants, variables, processes, actions, guards, and state changes, then produces a detailed model-generation plan using structured semantic prompts. A code generation LLM utilizing Retrieval Augmented Generation (RAG) converts the plan into CSP# code for the Process Analysis Toolkit (PAT) model checker. PAT verifies the generated model against user-specified requirements, and if verification fails, a repair loop uses counterexamples and violation traces to guide localized model edits. The verification-guided repair loop is necessary because many initially generated models compile but fail some requirements until counterexamples are used to correct them [26].

Also presenting a tool, Nivon et al. aim to reduce the barrier to formal modeling and model checking by using LLMs to translate NL system descriptions into verifiable formal models [27]. They present the tool GIVUP, which stands for GeneratIon and Verification of Underspecified Processes. GIVUP automatically generates Business Process Model and Notation (BPMN) models in the form of activity diagrams and an LTL property to verify from NL descriptions of a business process. The property is then verified via model checking and a diagnostic is returned if the property is violated. The framework translates BPMN processes into formal representations and LTL properties into Büchi automata to support model checking. GIVUP utilizes fine-tuned GPT-4o for BPMN activity diagram generation and LTL property generation. Notably, the authors found that fine-tuning GPT-4o on task-specific datasets yielded high accuracy and speed and indicate that the full GIVUP pipeline is practical for real-world process sizes [27].

Sultan and Apvrille developed TTool-AI, an LLM-based Model-Driven Engineering assistant integrated within the SysML modeling and formal verification toolkit TTool [28]. TTool-AI extends TTool with an LLM assistant that generates various SysML diagrams from NL system specifications. The TTool toolkit supports formal modeling, model verification (simulation and model checking), and code generation. TTool diagrams are expressed using formal semantics, enabling direct model checking of SysML models without translation into intermediate formalisms. Verification of Computation Tree Logic* (CTL*) properties is also supported [28].

On a separate note, Moezkarimi et al. utilize LLMs to translate a UML state diagram into a formal model, specifically a Rebeca model, that can then be model checked through the Afra model checker [29]. Rebeca is an actor-based modeling language for formally modeling concurrent reactive systems. Utilizing LLMs to translate UML models into formal representations for model checking can help ease the use of formal verification in the architecture and design phase of software development and allow developers to verify their designs before implementation. In their experiments, GPT-4 never generated formal models that compiled on the first attempt, as the LLM struggled with syntactic correctness and hallucination. The authors manually corrected formal models before performing model checking. Even with the compilation issues, the approach shows promise, as utilizing LLMs to translate UML diagrams into formal representations can significantly reduce the manual effort required for this task [29].

*2) Other Methods that Support Automating Model Checking*

Seven articles propose alternative methods to support automating model checking with LLMs [20, 21, 22, 30, 31, 32, 33]. Hasan et al. evaluate how correctly LLMs repair errors in Alloy specifications. Alloy specifications can then be used for downstream model checking [20]. The selected LLMs correct errors found within human-written specifications and are given varying levels of guidance through different prompting techniques and scenarios. Scenarios vary by whether the LLMs are provided with the location of the error in the specification, fix comments to help guide the fix, and assertion requirements. They found that the selected LLMs (GPT-3.5-Turbo, Gemini, PaLM/PaLM2, and Llama-3-70b) fall short when compared to existing techniques. Users therefore should not rely on them for repairing Alloy specifications. The LLMs performed well on syntactic repairs but struggled with semantic issues [20].

Wu and Lee investigate whether an LLM (specifically GPT-4) can effectively generate formal properties for UPPAAL timed automata given varying levels of contextual information [21]. They argue that for large and complex systems, humans may miss some critical properties because it is challenging to account for all critical properties in such systems. This task requires expertise, and humans may lack a comprehensive understanding of the whole system to identify all critical properties. The authors found that providing GPT-4 with a complete set of contextual information did not necessarily yield the best results. Specifically, they found that GPT-4 could generate formal properties based on a UPPAAL timed automata model without access to the full system requirements and vice versa. They recommend a collaboration between the developer and the LLM, where the developers first write a list of formal properties for a given system, and the LLM is used afterward to cover any blind spots and specify properties that may have been missed by the developer [21].

Moreira and Campos aim to lower the barrier to adopting model checking by enabling designers and domain experts (who lack formal verification expertise) to access NL explanations of model checking counterexamples [22]. They propose three alternative pathways for using LLMs to explain model checking counterexamples:

1) Creating a chatbot specialized in explaining counterexamples,
2) Using a general-purpose chatbot to generate explanations directly from the counterexample trace, and
3) Using a general-purpose chatbot to improve explanations already generated by an existing pattern-based approach.

The authors focus on preliminary experiments with pathways 2 and 3, providing the LLM with prompts containing the CTL property and the NuSMV counterexample, with varying levels of additional context. They used GPT-4o and LLaVa 1.5 throughout the experiments. The authors found that while LLMs offer flexibility in this task, they suffer from inconsistency, erroneous explanations in a considerable number of cases, and prohibitive computational cost for local models powerful enough to be useful. The overall quality of the results depends heavily on the model used; simpler models do not produce adequate results [22].

Separately, Yadavally et al. explore how LLMs can be used for the safe minimization of constraints [30]. Safe minimization involves finding the Minimum Unsatisfiable Subset (MUS) of a constraint system, where a MUS is an unsatisfiable subset of a constraint system such that removing any constraint from the set will make the set unsatisfiable. Constraints are used for model checking broadly, and this framework therefore supports minimizing the set of constraints used when model checking. The authors found that LLMs show significant promise in macro-reasoning for minimizing constraint systems [30].

Tangentially, Wu et al. combine LLMs and BMC to create a stable, training-data-free neuro-symbolic system for automatic loop invariant inference [31]. The authors propose LaM4Inv, which stands for LLM and Model checking for loop Invariant inference. LaM4Inv is a closed-loop framework that uses iterative prompting to incorporate feedback. The LLM is first prompted to generate a set of candidate loop invariants, which are each verified by an SMT solver. BMC is then used to evaluate individual predicates split from failed candidates. Predicates that hold across all checked program points are retained in a correct predicate set. The retained predicates are then reassembled into new candidate invariants and resubmitted to the SMT solver. Counterexamples and their violation types are then fed back into the LLM prompt for the next iteration. This loop repeats until a valid invariant is found or the resource budget is exhausted [31]. Although not directly about automating model checking, the authors present a noteworthy framework that combines LLMs and BMC to infer loop invariants. Such an approach could potentially be adapted to infer elements of a formal model or specification, with model checking used to obtain counterexamples that can then be used for refinement [31].

Pirzada et al. focus on improving BMC for C programs with unbounded or large-bound loops, where conventional loop unrolling often fails to prove correctness [33]. They propose replacing loops with verified loop invariants, using LLMs to generate candidate invariants. They then use a theorem prover to validate the generated candidate invariants. The motivation is to strengthen BMC's ability to prove safety without relying on loop unrolling, potentially increasing the number of programs that can be verified by BMC.

In model-based safety assessment, models used for formal safety analysis must accurately reflect the current textual system specification. In practice, however, specifications and design models often evolve in parallel, leading to inconsistencies in naming, structure, and detail—making manual comparison tedious and error-prone. Gonschorek et al. aim to automate the validation of Simulink design models against textual specification documents using LLMs [32]. They subsequently adapt validated models to incorporate safety requirements for formal analysis via model checking. The authors found that LLMs can successfully automate the extraction, mapping, and validation of architectural components and connections from engineering documents.

### *B. Formal Models for Downstream Formal Verification*

Seven articles propose methods that utilize LLMs to support downstream formal verification aside from only model checking. Corazza et al. utilize LLMs to generate formal models in the Quint modeling language for the formal verification of smart contracts used within blockchain systems [34]. Quint is based on TLA+. Experiments on GPT-4o found that the LLM generated correct Quint models within only a few iterations most of the time. Separately, Jha et al. combine inductive methods (LLM generation) with deductive methods (SMT-solving) to generate formal artifacts, formally verify them, and refine the formal artifact with the generated counterexample [35]. They utilize counterexample guided inductive synthesis to prompt the LLMs.

Harel et al. focus on methods to integrate chatbots into formal system modeling in a safe, structured, and controlled manner that reduces engineering effort without sacrificing correctness [36]. They propose leveraging scenario-based modeling as the target formalism because its compositional and human-readable structure facilitates both manual inspection and automated formal verification of LLM outputs via SMT-based analysis or model checking. On the other hand, Mao et al. aim to reduce the effort required to build symbolic models for security protocol verification [37]. Tools such as Tamarin and Proverif are powerful but difficult to use because users must manually translate NL protocol documents into precise formal models. The authors use LLMs to automatically generate trustworthy symbolic protocol models from NL protocol descriptions. They perform formal verification on generated SAPIC+, Tamarin, Proverif, and Deepsec formal models.

Wu et al.'s contribution is discussed in more detail in the previous section, yet their work contains applications to downstream formal verification outside of automating model checking [31]. The authors utilize SMT solving, specifically the Z3 engine, to verify candidate loop invariants in their combined LLM and BMC loop invariant inference framework. Their work in verifying invariants via SMT solving may be applied to formally verifying the correctness of programs that utilize loops, such as potentially unbounded loops.

Carvalho et al. investigate whether LLMs can help repair faulty formal specifications, specifically Event-B models [38]. Event-B models can then be used for theorem proving within the Rodin platform. The authors focus on iterative and small repairs rather than verifying an entire formal model at once. Experiments show that LLMs can effectively repair many small Event-B model faults when guided by carefully engineered prompts and formal verification feedback.

Requirements elicitation and initial modeling are among the most poorly supported activities in software development, and NL is often imprecise [39]. Klimek aims to bridge the gap between informal requirements and formal verification by proposing the Logic-oriented Requirements Engineering (LoRE+) formal framework [39]. LoRE+ uses LLMs to generate structured UML activity diagrams from NL scenarios, automatically translates those diagrams into formal logic specifications, performs deduction-based verification, and generates code skeletons from the activity diagrams, all within an integrated development environment.

### *C. Iterative and Repair-Based Methods*

This section discusses research that incorporates iterative or repair-based methods into LLM-generation frameworks. Twelve articles clearly incorporate iterative and repair-based methods, indicating a trend within the literature that developing such methods may increase the correctness of LLM generations for formal modeling.

Nour Eldin et al. propose a two-phase hybrid pipeline that utilizes a self-healing auto-correction mechanism in the first phase [40]. After the LLM extracts dependency relations between activities, the pipeline runs conventional graph-based algorithms to verify the structural integrity of the resulting dependency graph. Jha et al. co-synthesis code and SMV models for NuSMV symbolic model checking using LLMs [25]. When an error is discovered in either the code or the formal model via a counterexample from model checking, a syntax error, or a discovered compilation error, the LLM is iteratively prompted to correct the error in the respective artifact and update the remaining artifacts to reflect the change.

Jha et al. use counterexample guided inductive synthesis to prompt an LLM to generate formal artifacts [35]. An LLM is prompted to generate a well-structured output that is then checked within a formal verification environment. The counterexample from the formal verification step is included within the new prompt, and the LLM is prompted to generate the formal artifact again. The counterexample-guided iterative generation guides the LLM away from hallucinations and errors, bounding generations away from incorrect solutions [35]. Similarly, in the multi-stage LLM framework presented by Zuo et al., a model checker verifies the LLM-generated formal specifications [26]. If verification fails, a repair loop uses counterexamples and violation traces to guide localized model edits.

Harel et al. propose an iterative and LLM-agnostic method in which the LLM is given a preamble describing the formalism semantics, then prompted incrementally to generate individual scenario objects from NL requirements [36].

Outputs are independently verified via model checking and human code review. When errors are found, prompts are revised and the LLM is re-prompted to generate the correct model. Using an LLM, Mao et al. generate lambda expressions in an intermediate representation when generating symbolic security protocol models from NL documents [37]. These lambda expressions are then repaired using static analysis before proceeding with transformation into formal specifications.

Sultan and Apvrille present an extension to TTool called TTool-AI, which generates various SysML diagrams from NL system specifications [28]. The framework includes an iterative feedback loop to refine the generations. LLM responses are algorithmically analyzed for response structure errors and constraint errors, and queries are iteratively reformulated with detected errors until convergence, or a maximum iteration limit is reached. Similarly, Corazza et al. iteratively test and repair the LLM-generated model using prompts to guide the LLM toward the correct repair [34]. The prompts take into consideration the source of the error discovered (static, runtime, or semantic).

Zhang et al. propose a framework for the automatic generation of formal simulation models through fine-tuned LLMs [41]. They incorporate a validation and simulation framework for iterative syntax correction, human-in-the-loop refinement, and simulation-based verification. The framework by Gheorghita et al. supports iterative prompt-based model refinement through a conversational interface for correcting LLM-generated software diagrams [42]. On a similar note, Alshareef et al. use an iterative approach for generating activity diagrams with LLMs [43]. They provide iterative instructional prompts to generate and correct activity model instances. Lastly, Ahmed et al. incorporate a mechanism within their multi-stage diagram repair framework to filter out hallucinated issues from the LLM generation [44].

### *D. UML Model Generation*

Regarding UML model generation, three articles focus on sequence diagrams [44, 45, 46], three focus on domain/class models [47, 48, 49], and two explore the generation of multiple types of UML diagrams [42, 50]. UML models are semi-formal models used for the analysis and design of software systems. This research is included in the literature review because work on the automated generation of such diagrams is applicable to the automated generation of formal models for model checking.

#### *1) Sequence Diagrams*

Ferrari et al. investigate whether an LLM can generate UML sequence diagrams from NL requirements [45]. The authors use GPT-3.5 to translate NL software requirements into UML sequence diagrams. The generated diagrams are semi-formal behavioral models that represent interactions among system actors and components. The prompting strategy follows the "visualization generator" pattern: the LLM is asked to generate a sequence diagram from a list of requirements so that it can be visualized in PlantText [45].

Similarly, Jahan et al. study the effectiveness of GPT-3.5 in generating UML sequence diagrams from NL user stories [46]. They prompt the LLM with individual user stories in a zero-shot setting to generate a corresponding UML sequence diagram in PlantUML script format. They found that GPT-3.5 consistently generated sequence diagrams without syntax errors. However, for short and simple user stories, GPT-3.5 tended to overelaborate, adding contextual elements not explicitly stated in the user story and sometimes producing overly complex diagrams [46]. Lastly, Ahmed et al. developed the Model Correctness evaluation Tool (MCeT) that uses LLMs to evaluate whether a sequence diagram is accurate and complete with respect to requirements text [44]. MCeT provides NL explanations to the developers on discovered accuracy and completeness issues.

#### *2) Domain/Class Models*

Chen et al. evaluate the extent to which LLMs can fully automate domain modeling directly from an NL problem description and without any human interaction or supervised training [47]. They evaluate both GPT-3.5 and GPT-4. The authors formulate domain modeling as a text generation problem. Given an NL problem description, the LLM generates a textual representation of a domain model containing enumerations, classes (with attributes), and relationships (associations, compositions, and generalizations) in a custom Extended Backus-Naur Form format [47]. On a similar note, Siala and Lano focus on automatically generating UML class diagrams and Object Constraint Language (OCL) specifications from code using fine-tuned LLMs [48]. Their framework abstracts class diagrams and OCL specifications from both Java and Python source code.

Cámara et al. explore the capabilities of ChatGPT (stable release of February 2023) to perform modeling tasks and assist modelers [49]. The authors use ChatGPT as a modeling assistant to generate UML class diagrams from NL prompts. They specifically investigate the LLM's ability to produce syntactically and semantically correct models in various textual UML notations (e.g. PlantUML); handle modeling concepts such as association classes, multiple inheritance, enumerations, and OCL constraints; and perform cross-modeling language translation. Notably, the authors found that the LLM performed better when asked to build a small initial model and then iteratively add detail to the model. They also found that syntactic correctness was generally adequate, though semantic correctness was weak [49].

#### *3) Multiple Diagram Types*

Gheorghita et al. propose DiagramAI, a tool for AI-driven generation and editing of software diagrams (use case, class, architectural, sequence, and C4 context diagrams) [42]. LLMs are prompted to produce either a direct model output or an intermediate representation that is then converted to the output model via a rule-based post-processing pipeline. Experiments indicate that LLMs show significant potential for assisting in diagram generation but are not yet fully reliable for autonomous diagram construction when precise syntax and structural correctness are required [42].

Nguyen et al. address two research problems: (1) how to automatically generate a comprehensive UML dataset that captures both structural and behavioral aspects in a scalable and semantically coherent manner, and (2) how to systematically validate the correctness, consistency, and usefulness of automatically generated UML datasets [50]. The authors focus on the automated generation and validation of UML diagrams expressed as PlantUML code, spanning six diagram types: class, object, component, use case, sequence, and state diagrams

### *E. SysML Model Generation*

Manual SysML behavioral modeling is slow, error-prone, and costly, limiting the practical adoption of model-driven development [51]. There is therefore growing interest in leveraging LLMs to automate aspects of SysML behavioral modeling to ease its practical use. Work on automating SysML modeling, like UML modeling discussed in the previous section, is applicable to formal model generation and is therefore included in this literature review.

As described in section IV.A, Sultan and Apvrille present TTool-AI, an LLM extension to TTool for automating SysML modeling [28]. TTool-AI supports the automatic generation of use case diagrams, block diagrams, and state machines. The authors evaluate the tool on GPT-3.5-Turbo and GPT-4o. The overall purpose of TTool-AI is to provide strong first drafts of models that the user can then refine and improve, not to replace the human modeler. They found that TTool-AI generates high-quality SysML diagrams comparable to or better than those of Master-level students, with significantly reduced generation times.

Similarly, Wang et al. evaluate how effective LLMs are at generating three types of SysML behavioral models: state machines, activity diagrams, and sequence diagrams [51]. They propose a two-phase LLM framework. In the first phase, a structured prompt template is used to generate the corresponding SysML models. In the second phase, model checking rules derived from hallucinations from the first phase are injected into the prompt's error slot, resulting in LLM regeneration in a new session.

Alshareef et al. examine how LLMs can assist in constructing simulation models, specifically activity and flow-based diagrams applicable to SysML, UML, and systems engineering [43]. They demonstrate methods to incorporate LLM output into the modeling process as a starting point for further elaboration by modelers. Lastly, Xu et al. propose an LLM-enhanced SysML framework (LLM-SysML) that utilizes LLMs to improve the accuracy of the transformation from NL requirements to formal SysML models [52]. Three SysML diagram types are produced: use case diagrams, sequence diagrams, and requirements diagrams. The authors propose a four-step framework which includes (i) requirements collection and analysis, (ii) LLM-assisted element abstraction, (iii) SysML model implementation, and (iv) performance assessment and iterative updating.

### *F. Generating Models from Code*

Only two articles in this literature review focus specifically on generating (semi) formal artifacts from code [48, 53]. Boronat and Mustafa address the challenge of recovering high-level abstractions (domain models) from low-level source code by integrating LLMs with traditional Model-Driven Reverse Engineering (MDRE) techniques [53]. They introduce MDRE-LLM, which combines LLM capabilities with RAG to automate the recovery of domain models from Java codebases. Siala and Lano also support MDRE by using LLMs to abstract accurate and comprehensible UML and OCL representations from Java and Python source code [48]. They propose LLM4Models, which uses fine-tuned LLMs to automatically generate UML classes and OCL specifications from code. Both articles present experiments that demonstrate the effectiveness and feasibility of LLMs for abstracting models and specifications from code.

## V. Discussion

This section answers RQ2 by discussing common themes found within the literature. Specifically, this section discusses the performance seen across the literature on LLM-aided formal modeling, which LLMs generally showed strong or weak performance, common prompting techniques across sources, developer tools that have been rigorously evaluated, and common ideas discussed in the literature.

### *A. Performance of LLMs on Formal and Semi-Formal Modeling*

Of the 34 reviewed articles, 16 report overall strong and/or promising results for utilizing LLMs for the automated generation of formal and semi-formal models: Saccon et al. [23], Ramani et al. [24], Zuo et al. [26], Nivon et al. [27], Sultan and Apvrille [28], Yadavally et al. [30], Wu et al. [31], Gonschorek et al. [32], Pirzada et al. [33], Corazza et al. [34], Carvalho et al. [38], Klimek [39], Zhang et al. [41], Jahan et al. [46], Siala and Lano [48], and Boronat and Mustafa [53].

On the other hand, five articles report relatively weak results on this task: Hasan et al. [20], Moreira and Campos [22], Moezkarimi et al. [29], Gheorghita et al. [42], and Ferrari et al. [45]. All other articles reported results somewhere in between, where the LLMs show notable results that were neither significantly strong nor weak.

Table IV presents the LLMs used within the literature and lists which articles found strong, weak, or neutral (neither strong nor weak) performance for each LLM. A variety of LLMs were selected for experiments. Several articles did not precisely state which LLM was used and mentioned generic identifiers for qualitative experiments only. These results are not included in Table IV.

TABLE IV
LLM PERFORMANCE IN REVIEWED ARTICLES

| LLM | Strong | Weak | Neutral |
|---|---|---|---|
| BERT | [41] | | |
| CodeQwen1.5-7B | [41] | | |
| Mistral-7B-v0.3 | [48] | | |
| Zephyr-7B | | [42] | |
| LlaVa 1.5 | | [22] | |
| Microsoft Phi-3 | | [42] | |
| Llama-3-8B | [31] | | |
| Llama-3-70b | | [20] | |
| Llama-3.1 405B Results | | [25] | |
| Meta Llama 3.1 | | | [51] |
| Meta Llama-3.2-11B-Vision-Instruct | | [42] | [50] |
| DeepSeek R1 | | | [50] |
| DeepSeek v3 | | | [51] |
| Grok 4 | | | [32] |
| PaLM/PaLM 2 | | [20] | [43] |
| Kimi Moonshot-v1 | | | [51] |
| Claude 3 Haiku | | | [51] |
| Claude-3.5 Sonnet | [25, 30] | | |
| Claude-3.7 Sonnet | [26] | | |
| OpenAI's o3-mini | [26, 38] | | |
| GPT-O3 | [32] | | |
| GPT-3.5 | [46] | [30, 45] | [47] |
| GPT-3.5-Turbo | [28, 31, 33] | [20, 35] | [37] |
| GPT-4 | [30, 31] | [29] | [21, 30, 35, 37, 43, 47] |
| GPT-4o | [23, 27, 28, 34, 39] | [22, 24] | [25, 37, 40, 51] |
| GPT-4-Turbo | [31] | | [40, 51] |
| GPT-4o-mini | [53] | | [44] |
| GPT-5 | [24] | | |
| GPT-5-mini | [23] | | |

### *B. Common Prompting Techniques*

A variety of prompting techniques are used in the literature for automating formal and semi-formal modeling. These techniques, alongside the articles that employed them, are listed in Table V. Not all articles detailed their prompting strategies. Additionally, only techniques utilized by at least two articles are listed in Table V. Incremental approaches refer to prompting techniques where the final result is built incrementally using multiple prompts.

### *C. Developer Tools*

Several articles present user-facing developer tools for automated modeling with LLMs. These tools include:

1) PAT-Agent for the automated generation of CSP# specifications for model checking within the PAT model checker [26].
2) GIVUP for the automated generation of BPMN models in the form of activity diagrams and formal properties for formal verification [27].
3) TTool-AI for automated SysML modeling. This tool supports the generation of use case, block, and state machine diagrams [28].
4) DiagramAI, which supports the automated generation of use case, class, architectural, sequence, and C4 context diagrams [42].

TABLE V
COMMON PROMPTING TECHNIQUES

| Prompting Technique | Source |
|---|---|
| Chain-of-Thought [54] | [47, 30, 33, 44, 32, 50] |
| Zero-shot [54] | [46, 47, 20, 48] |
| One-shot [54] | [24, 25, 43, 39, 26] |
| Few-shot [54] | [40, 47, 23, 20, 21, 29, 41, 37, 44, 34] |
| Incremental approaches | [23, 36] |
| Role-based [55] | [40, 20, 51, 26] |

### *D. Common Themes and Ideas*

This section presents several common themes and ideas presented across the literature.

#### *1) Co-Generation of the Model and Formal Property*

Three articles propose generating both the formal model and formal property together for formal verification [23, 25, 27]. This area warrants further study, as it may increase the accessibility of model checking if formal property generation is also automated alongside formal model generation.

#### *2) Human Involvement throughout the Model Generation Process*

Human involvement throughout the model generation process would allow for early validation and correction of formal artifacts. The argument is that involving the developer often and early may save time compared to verifying and validating the final model and then regenerating if the model is inadequate. Mistakes can be caught early when they are both easy to understand and correct and before the final model is generated [39, 40].

#### *3) Collaboration Rather than Replacement*

Humans and LLMs should collaborate to develop a model or formal artifact, rather than the LLM replacing the human modeler. For example, an LLM can generate preliminary results that are then elaborated upon by the human developer. On the other hand, the developer can first develop a model or formal properties that are then iteratively improved by an LLM [21, 28].

#### *4) LLMs Extract Modeling Elements Only*

Rather than holistically generating the formal artifact from input requirements or code, the LLM would instead extract elements of the formal model, such as processes, entities, and relationships. A deterministic algorithm can then be employed to construct the model from these components. Focusing on extracting elements of the model using an LLM, then deterministically generating the model, can lead to more accurate and understandable models. This process also supports human validation as the steps are easier to verify [39, 40, 42].

#### *5) Syntactic Repairs vs. Semantic Correctness*

Several experiments found that LLMs generally perform well on syntactic repairs and generations, but struggle with semantic correctness [20, 43, 45, 49, 51]. This finding especially limits the practical use of LLMs for formal artifact generation and is a significant area for future work.

#### 6) LLMs Require Strict Guidance or Intermediate Representations

LLMs require strict guidance or intermediate representations during the generation process to generate correct formal artifacts. The direct approach to LLM prompting is unreliable as it may lead to hallucinations or the difficulty for the LLM to focus on the entire context provided by the user. Therefore, more structured pipelines or protocols are necessary to improve the LLM's performance. In other words, the direct or "holistic" approach of providing NL to an LLM and generating formal models is less effective than having an intermediate step between the LLM receiving the NL requirements and generating the formal model [23, 26, 27, 33, 36, 37, 40, 44].

#### 7) Prompt Optimization with LLMs

Using an LLM at the beginning of the artifact generation pipeline for prompt optimization may result in more successful prompts. The optimized prompt would then be used by a downstream LLM when generating the artifact [39, 40].

## VI. Gaps, Challenges, and Areas for Future Work

This section answers RQ3 by discussing the gaps, challenges, and areas for future work in the research area.

### A. Gaps and Challenges

The reviewed literature reveals a strong bias toward generating formal and semi-formal artifacts from NL descriptions rather than from existing source code. Only two articles in this literature review focus specifically on recovering formal or semi-formal models from code, despite the knowledge that many real-world software systems already exist in code form without corresponding formal specifications [48, 53]. This imbalance suggests that reverse-engineering formal models from legacy or existing codebases remains a largely unexplored problem, even though such capabilities could meaningfully reduce the verification burden for systems that were never formally specified during initial development. Another concern is that the LLMs evaluated throughout the literature are notably outdated relative to the current state of the art. The most recent models employed were GPT-5 [24] and GPT-5-mini [23], each used in only a single article, while the majority of studies rely on earlier models such as GPT-3.5, GPT-4, and GPT-4o. Given the rapid pace of LLM development, conclusions drawn from these older models regarding LLM capability and reliability for formal modeling tasks may no longer accurately reflect the performance achievable with current systems.

A second challenge concerns the scale and rigor of evaluation across the literature. Most proposed frameworks are evaluated on small, hand-selected datasets or case studies, and several articles report primarily qualitative results rather than large-scale quantitative benchmarks [36, 39, 43, 52]. Very few studies evaluate their approaches on large or industrially representative systems [27], which limits confidence in the real-world applicability of these techniques. This limitation is compounded by the fact that evaluating LLM-generated formal artifacts at scale remains inherently difficult; many authors resort to manual inspection of generated models to assess correctness, which restricts the number of examples that can feasibly be evaluated and introduces subjectivity into the reported results. Although some frameworks mitigate this issue by embedding formal verification directly into an iterative generation-and-correction loop [26, 35], such automated validation is not yet standard across the literature, and manual evaluation continues to be a bottleneck to rigorous, reproducible assessment.

Another gap is the relatively limited attention paid to automating model checking itself, as opposed to the broader task of formal model generation for software engineering activities more generally. While seven articles address the automated generation of formal models specifically intended for model checking [23, 24, 25, 26, 27, 28, 29], and a similar number explore adjacent tasks such as specification repair, property generation, and counterexample explanation [20, 21, 22, 30, 31, 32, 33], this remains a comparatively small subset of the literature. Compounding this gap is the inherent nondeterminism of LLMs, which poses a persistent challenge to the reproducibility of any model checking or formal modeling pipeline. Because repeated queries with identical prompts can yield inconsistent outputs (i.e. some correct and others flawed), reliably reproducing reported results, or guaranteeing consistent behavior in a deployed tool, remains an open problem that few articles address directly [38, 42, 44].

Finally, a recurring and significant challenge throughout the literature is the disparity between syntactic and semantic correctness in LLM-generated artifacts. Multiple studies report that LLMs are generally capable of producing syntactically valid formal models or specifications, yet frequently fail to capture the intended semantics of the system being modeled [20, 43, 45, 49, 51]. In several cases, semantic evaluation is largely omitted, with authors focusing specifically on whether the generated artifact compiles or parses correctly rather than whether it faithfully represents the underlying requirements [24, 25, 42, 43]. This distinction is particularly consequential in the context of model checking and formal verification, where a syntactically valid but semantically incorrect model leads to unsound verification results, potentially giving developers false confidence in the correctness of their systems

### B. Areas for Future Work

A promising direction for future research is the continued exploration of automating model checking with LLMs specifically. Co-synthesizing the formal model and its corresponding formal property is a particular area for further exploration. Only three articles in this review generate both artifacts together within a single framework [24, 25, 27]. Automating property specification alongside model generation may further lower the barrier to adopting model checking for developers without formal verification expertise. Pursuing this direction introduces its own challenges, however, as ensuring the joint correctness of both the model and the property likely requires human involvement at multiple stages of the generation pipeline.

Another area for future work concerns strengthening the correctness guarantees of LLM-generated artifacts, both syntactically and semantically. Given that several studies report LLMs performing well on syntactic tasks while struggling with semantic fidelity, future work should prioritize the development of evaluation methods and generation pipelines that explicitly target semantic correctness rather than treating successful compilation or parsing as a metric for overall quality. One promising strategy is the integration of formal verification directly into the generation loop, where model checking or theorem proving are used not only to validate a final artifact but to iteratively guide its construction and correction, thereby reducing the manual effort required to verify LLM output [26, 35]. In parallel, future research should investigate methods to mitigate the effects of LLM nondeterminism on reliability and reproducibility, for instance through ensembling multiple generations or verification-guided selection among candidate outputs, so that practitioners can have greater confidence that a given framework will behave consistently across repeated invocations.

Additionally, the scalability and currency of existing approaches warrant further investigation. Because most frameworks in this review have been evaluated only on relatively small or simplified systems, future work should explore methods for applying LLM-based formal modeling to large and complex real-world systems. Future research might focus on pipelines that decompose such systems into smaller, more tractable components that can be modeled and verified individually before being composed into a complete artifact. Relatedly, given that the LLMs evaluated throughout the literature are largely outdated, there is a clear need to re-evaluate existing frameworks using state-of-the-art models. It is plausible that approaches previously reported as yielding weak or neutral results may perform substantially better when paired with more capable and recent LLMs. Such re-evaluation would help clarify whether observed limitations stem from the underlying framework design or from the capabilities of the specific LLM employed at the time of study. Future work might also further investigate whether using an LLM to optimize the prompt supplied to a downstream generation LLM meaningfully improves output quality [39, 40].

More broadly, techniques developed for adjacent tasks throughout this review, such as UML and SysML model generation, iterative repair frameworks, and code-to-model recovery, may be adaptable to the model checking context specifically. Investigating how these approaches transfer to the generation of formal models intended for model checking, rather than for other downstream verification or design purposes, represents a substantial and open avenue for future work.

Lastly, future work should extend beyond purely technical framework development to consider the human and organizational context in which these tools would be deployed. User studies examining the ease of use, learnability, and perceived trustworthiness of LLM-based formal modeling tools are necessary to demonstrate their practical applicability in industrial settings, particularly for practitioners who lack formal methods expertise. Such studies would also help clarify the appropriate division of labor between human developers and LLMs, building on the collaborative rather than replacement-oriented paradigm favored in the literature [21, 28]. Separately, an underexplored but potentially valuable application of LLMs in this space is their use in educational settings for teaching formal modeling, where an LLM familiar with a target formalism could assist students in learning to construct and reason about formal specifications [49, 53].

## VII. Related Work

This work is the first literature review focusing specifically on the development of LLM-based methods for the automated generation of formal models for model checking and formal verification. Although some reviews discuss software modeling [55, 56, 57], there is a clear gap in the literature in regard to reviewing formal modeling using LLMs. The most closely related work is by Fresemann et al., which reviews literature focusing on using LLMs to support modeling tasks, especially semi-formal modeling languages such as SysML [55]. However, their review is not systematic, and they do not focus specifically on formal modeling and formal verification.

Adjacently, Wang et al., Tasarsu et al., and Wang et al. provide literature reviews on LLMs applied to software testing activities [58, 59, 60]. Wang et al. specifically discuss integrating LLMs with formal methods to improve the coverage and redundancy of test case generation [58]. On the other hand, Beg et al., Zadenoori et al., Ciano et al., and Hemmat et al., review literature applying LLMs to requirements engineering activities [55, 56, 60, 61]. Beg et al., specifically survey the use of LLMs for writing formal specifications from software requirements [61]. Zadenoori briefly covers articles focusing on utilizing LLMs for requirements modeling, e.g. generating UML diagrams [55].

Zhang et al., Liu et al., Hou et al., and Fan et al. provide literature reviews focused on LLMs applied to software engineering broadly [63, 64, 65, 66]. Regarding formal methods, Hou et al. discuss specification formalization and formal verification briefly [65]. However, these articles do not focus on automated formal modeling, further emphasizing the gap in the literature on this topic.

Separately, Dolcetti and Lotti, Weng et al., Li et al., and Li et al., review other tangentially related topics [67, 68, 69, 70]. Dolcetti and Lotti review the usage of LLMs for software verification focusing on two main topics: (i) using LLMs as verification tools, and (ii) verifying code produced by LLMs [67]. They briefly discuss combining LLMs with formal verification tools, though this is not a main focus. Weng et al. present a survey on the autoformalization of informal mathematical propositions into verifiable formal representations [68]. This task supports automated theorem proving in mathematics. They include research utilizing LLMs for this task. Li et al. systematically review literature focusing on LLMs for the testing and verification of autonomous intelligent systems [69]. They cover formal verification alongside test scenario generation, vulnerability detection, and real-time monitoring. Lastly, Li et al. present a systematic literature review on using LLMs in various stages of ontology development [70].

## VIII. Conclusion

This paper presents an SLR of 34 articles examining how LLMs are being applied to automate model checking and the generation of formal and semi-formal models within software engineering. The review found that current research clusters around six overlapping trends: automating the generation of formal models for model checking, supporting downstream formal verification tasks more broadly, incorporating iterative and repair-based mechanisms to correct LLM output, generating semi-formal UML and SysML artifacts that share transferable techniques with formal model generation, and generating models from code. Across these trends, a consistent pattern emerged in which LLMs demonstrate reasonably strong syntactic capabilities but comparatively weak semantic fidelity, motivating the widespread adoption of counterexample-guided repair loops, intermediate representations, and human-in-the-loop validation as mitigating strategies. Notably, the use of LLMs specifically to automate model checking, as opposed to the broader and more studied task of formal or semi-formal model generation, remains a largely unexplored subarea of this literature. Only a small number of the reviewed articles directly address automated model checking, and this SLR found little research dedicated to this narrower topic prior to 2024. Nevertheless, interest in this area has grown steadily, with a marked increase in work published in 2025 and into early 2026. This suggests that the intersection of LLMs and model checking is only beginning to attract sustained attention from the research community.

Despite the immaturity of this body of work, the reviewed literature suggests that LLMs hold substantial promise for reducing the manual effort and specialized expertise historically required to construct, repair, and interpret formal models. Continued progress in this direction has the potential to meaningfully ease the use of formal verification and to increase the accessibility of model checking to practitioners who lack formal methods training. Realizing this potential, however, will require the field to address the gaps identified in this review, including the need for evaluation on larger and more representative systems, re-evaluation of existing frameworks using state-of-the-art LLMs, greater emphasis on semantic rather than purely syntactic correctness, and a deeper investigation into the joint synthesis of formal models and their associated properties. Given the sparse and largely unexplored state of research specifically targeting LLM-based automation of model checking, this SLR is intended to serve as a foundation and roadmap for researchers seeking to contribute to this emerging area. The broader synthesis of adjacent formal and semi-formal modeling work offers a set of techniques that may be adapted and extended toward that end.